\documentclass[twocolumn,showpacs,prd]{revtex4}
\usepackage{mathrsfs,bm,multirow}
\usepackage{longtable,lscape}
\usepackage{txfonts}
\usepackage{amssymb}
\usepackage{indentfirst}
\usepackage{graphicx,,booktabs}
\usepackage{color}

\begin{document}

\title{Reproducing the $Z_c(3900)$ structure through the initial-single-pion-emission mechanism}
\author{Dian-Yong Chen$^{1,3}$}
\email{chendy@impcas.ac.cn}
\author{Xiang Liu$^{1,2}$\footnote{Corresponding author}}\email{xiangliu@lzu.edu.cn}
\author{Takayuki Matsuki$^4$}
\email{matsuki@tokyo-kasei.ac.jp}
\affiliation{$^1$Research Center for Hadron and CSR Physics,
Lanzhou University $\&$ Institute of Modern Physics of CAS,
Lanzhou 730000, China\\
$^2$School of Physical Science and Technology, Lanzhou University,
Lanzhou 730000, China\\
$^3$Nuclear Theory Group, Institute of Modern Physics, Chinese
Academy of Sciences, Lanzhou 730000, China\\
$^4$Tokyo Kasei University, 1-18-1 Kaga, Itabashi, Tokyo 173-8602,
Japan}

\begin{abstract}

Being stimulated by the recent BESIII observation of a charged charmoniumlike structure $Z_c(3900)$, in this work we study the distributions of the $J/\psi\pi^\pm$ and $\pi^+\pi^-$ invariant mass spectra of the $Y(4260)\to \pi^+\pi^- J/\psi$ decay by the initial-single-pion-emission mechanism, where the interference effects of the ISPE mechanism with two other decay modes are also taken into account. The obtained $d\Gamma(Y(4260)\to \pi^+\pi^- J/\psi)/dm_{J/\psi \pi^\pm}$ and $d\Gamma(Y(4260)\to \pi^+\pi^- J/\psi)/dm_{\pi^+ \pi^-}$ marvelously agree with the BESIII data to reproduce the $Z_c(3900)$ structure.

\end{abstract}
 \maketitle


Very recently the BESIII Collaboration announced the observation of
a charged charmoniumlike structure $Z_c(3900)$, which is near the
$D\bar{D}^*$ threshold and comes from the analysis of the
$J/\psi\pi^\pm$ invariant mass spectra in $e^+e^-\to \pi^+\pi^-
J/\psi$ at $\sqrt{s}=4260$ MeV \cite{BESnew}. Before this
interesting and important experimental observation, there were
some theoretical predictions of charged charmoniumlike structures
near $D\bar{D}^*$ and $D^*\bar{D}^*$ thresholds
\cite{Chen:2011xk,Chen:2012yr,Sun:2011uh}, which were mentioned in
the BESIII paper on $Z_c(3900)$ \cite{BESnew}.

The BESIII's result on $Z_c(3900)$ has also inspired extensive
discussions of its properties
\cite{Wang:2013cya,Guo:2013sya,Faccini:2013lda,
Voloshin:2013dpa,Karliner:2013dqa,Mahajan:2013qja,
Cui:2013yva,Wilbring:2013cha} and the prediction of a charged
charmoniumlike structure with hidden-charm and open-strange
\cite{Chen:2013wca}. What is more important is that the Belle
Collaboration confirmed $Z_c(3900)$ in $e^+e^-\to \pi^+\pi^- J/\psi
$ at $\sqrt{s}=4260$ MeV \cite{Liu:2013dau}. Later, the former
CLEO-c group analyzed data of their own and observed the charged
$Z_c(3900)$ structure in their analysis of 586 pb$^{-1}$ data taken
by their detector at $\psi(4160)$ \cite{Xiao:2013iha}, which is highly
consistent with the BESIII's observation \cite{BESnew}. In addition,
the neutral $Z_c(3900)$ was also released at a 3$\sigma$
significance \cite{Xiao:2013iha}.

Among many theoretical papers
\cite{Chen:2011xk,Chen:2012yr,Sun:2011uh} on the predictions of
charged charmoniumlike structures cited in BESIII's paper
\cite{BESnew}, Ref. \cite{Chen:2011xk} has given the predictions of
enhancement structures near $D\bar{D}^*$ and $D^*\bar{D}^*$
thresholds in the corresponding $J/\psi\pi^+$, $\psi(2S)\pi^+$, and
$h_c(1P)\pi^+$ invariant mass spectra, where $\psi(4040)$,
$\psi(4160)$, $\psi(4415)$, and $Y(4260)$ decay into $\pi^+\pi^-
J/\psi$, $\pi^+\pi^- \psi(2S)$, and $\pi^+\pi^- h_c(1P)$ via the
initial-single-pion-emission (ISPE) mechanism \cite{Chen:2011pv},
which was first proposed to understand the Belle's observations of
two charged bottomonium-like structures $Z_b(10610)$ and
$Z_b(10650)$ appearing in the hidden-bottom dipion decays of
$\Upsilon(5S)$ \cite{Collaboration:2011gja}. Later, the ISPE
mechanism was applied to study the hidden-bottom dipion decays of
$\Upsilon(11020)$, where the charged bottomonium-like structures
near the $B\bar{B}^*$ and $B^*\bar{B}^*$ thresholds were predicted
\cite{Chen:2011pu}. In addition, two charged strangeonium-like
structures observable in the $Y(2175) \to \pi^{+} \pi^{-}
\phi(1020)$ process were predicted in Ref. \cite{Chen:2011cj} using
the ISPE. These abundant theoretical results predicted by the ISPE
mechanism \cite{Chen:2011xk,Chen:2011pu,Chen:2011cj} can be further
tested in experiment.

Having more experimental information on the distribution of the
$J/\psi\pi^\pm$ and $\pi^+\pi^-$ invariant mass spectra of
$Y(4260)\to \pi^+\pi^- J/\psi $
\cite{BESnew,Liu:2013dau,Xiao:2013iha}, we can perform an
in-depth study of $Y(4260)\to \pi^+\pi^- J/\psi$ through the ISPE
mechanism to get deeper understanding of a $Z_c(3900)$ structure
discovered in the $J/\psi\pi^\pm$ invariant mass spectrum. Along
this road, in this paper we will fit our model based on the ISPE
mechanism to the experimental data of the
$J/\psi\pi^\pm$ and $\pi^+\pi^-$ invariant mass spectra of
$Y(4260)\to \pi^+\pi^- J/\psi $.
In this study, we would like to check whether the $Z_c(3900)$ structure can
be reproduced through the ISPE mechanism. This will be helpful to
deeply understand how to generate $Z_c(3900)$,
which makes our study presented here an intriguing research work.



\begin{figure*}[htbp]
\centering%
\begin{tabular}{c}
\scalebox{0.69}{\includegraphics{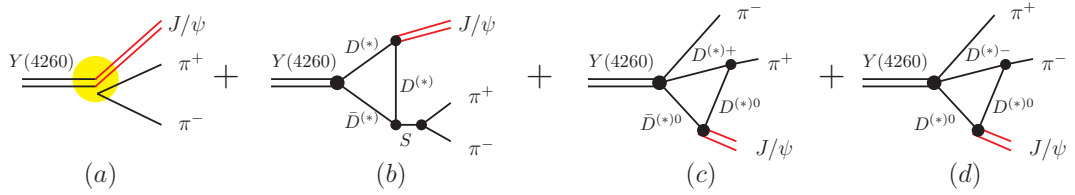}}%
\end{tabular}
\caption{(Color online.) The typical diagrams depicting $Y(4260) \to
J/\psi\pi^+\pi^-$ decay. Here, (a) denotes
$Y(4260)$ direct decay into $J/\psi\pi^+\pi^-$; Fig.
(b) describes the intermediate hadronic loop
contribution to $Y(4260)\to J/\psi\pi^+\pi^-$. Fig.
(c) and (d) are from the ISPE mechanism.
\label{Fig:ISPEDia}}
\end{figure*}

As for the hidden-charm dipion decay discussed here,
$$Y(4260)(p_0)\to \pi^+(p_1)\pi^-(p_2) J/\psi(p_3),$$
there exist different decay mechanisms as shown in Fig. \ref{Fig:ISPEDia}, which play
an important role in connecting the initial state $Y(4260)$ and final state $\pi^+\pi^- J/\psi$.
Among these decay mechanisms, the first one is due to the $Y(4260)$ direct decay into  $\pi^+\pi^- J/\psi$ (see Fig. \ref{Fig:ISPEDia} (a)), where $Y(4260)\to \pi^+\pi^- J/\psi$ occurs without the intermediate state contribution and mainly provides the background for the distributions of the $\pi^+\pi^-$ and $J/\psi\pi^\pm$ invariant mass spectra. The intermediate state contributions come from $\sigma(600)$, $f_0(980)$, and the triangle hadronic loops composed of charmed mesons.

By the effective Lagrangian approach, the decay amplitude of the direct decay of  $Y(4260)\to \pi^+\pi^- J/\psi$ can be written as
\begin{eqnarray}
&&\mathcal{M}_{\mathrm{Direct}} [Y(4260)\to \pi^+\pi^- J/\psi]\nonumber \\&&=
{{F}\over{f_\pi^2}}\epsilon_{Y(4260)}\cdot
\epsilon_{J/\psi}\Big\{\Big[q^2-\kappa (\Delta
M)^2\Big(1+\frac{2m^2_\pi}{q^2}\Big)\Big]_{\mathrm{S-wave}}\nonumber\\&&
\quad+\Big[\frac{3}{2}\kappa\big((\Delta M)^2-q^2\big)
\Big(1-\frac{4m_\pi^2}{q^2}\Big)
\Big(\cos\theta^2-\frac{1}{3}\Big)\Big]_{\mathrm{D-wave}}\Big\},\label{direct}
\end{eqnarray}
which was once constructed by Novikov and Shifman to study
$\psi^\prime\to \pi^+\pi^- J/\psi$ transition \cite{Novikov:1980fa}.
In Eq. (\ref{direct}), the subscripts S-wave and D-wave represent
the corresponding amplitudes from S-wave and D-wave contributions,
respectively. $\Delta M= m_{Y(4260)}-m_{J/\psi}$ is the mass
difference between $Y(4260)$ and $J/\psi$. $q^2=(p_1+p_2)^2=m_{\pi^+
\pi^-}^2$ denotes the invariant mass of $\pi^+ \pi^-$ while $\theta$
is the angle between $Y(4260)$ and $\pi^-$ in the $\pi^+ \pi^-$ rest
frame. In addition, the pion mass and decay constants are taken as
$m_{\pi^\pm}= 139$ MeV and $f_\pi =130$ MeV, respectively.
{Here, the amplitude of direct contribution is
constructed to study the dipion transitions between spin "$1^{++}$" particles.
However, the Lorentz structure of the vertex reflects only the quantum numbers
of the involved particles. Since the $Y(4260)$ and $\psi^\prime$
have the same quantum numbers, we can employ the amplitude proposed in Ref. \cite{Novikov:1980fa} to
describe the direct contribution of $Y(4260) \to J/\psi \pi^+
\pi^-$}. In the direct decay process, two free parameters, $F$ and
$\kappa$, are introduced, which can be determined by fitting to the
experimental data.

The second mechanism in the $Y(4260) \to J/\psi \pi^+ \pi^-$ decay
is due to intermediate $S=\sigma(600)$ and $f_0(980)$ contributions,
where the hadronic loop constructed by the charmed mesons is the
bridge to connect $Y(4260)$ with $J/\psi S$ as described by Fig.
\ref{Fig:ISPEDia} (b). In this work, we parametrize the decay
amplitude corresponding to Fig. \ref{Fig:ISPEDia} (b) as
\begin{eqnarray}
&&\mathcal{M}_{\mathcal{S}}  [Y(4260)\to \pi^+\pi^- J/\psi]\nonumber
\\&&= \epsilon_{Y(4260)}^\mu \epsilon_{J/\psi}^\nu\; g_{\mu \nu}\;
\frac{F_{\mathcal{S}}}{(p_1+p_2)^2-m_{\mathcal{S}}^2 +
im_{\mathcal{S}} \Gamma_{\mathcal{S}}},\label{second}
\end{eqnarray}
where a subscript $\mathcal{S}$ can be taken as $\sigma$ or
$f_0(980)$ and $m_{\mathcal{S}}$ and $\Gamma_{\mathcal{S}}$ are the
resonance parameters of $\sigma(600)$ and $f_0(980)$, respectively. Since the
total decay width and mass of $\sigma(600)$ are of the same order,
the minimal width approximation is not valid. Here, we adopt the
momentum dependent decay width in the propagator of $\sigma(600)$,
i.e.,
\begin{eqnarray}
\Gamma_\sigma(m_{\pi ^+\pi^-}) =\Gamma_{\sigma} \frac{m_{\sigma}}{m_{\pi^+\pi^-}}
\frac{|\vec{p}(m_{\pi^+\pi^-})|}{|\vec{p}(m_{\sigma})|},
\end{eqnarray}
where $|\vec{p}(m_{\pi^+\pi^-})|=\sqrt{m_{\pi^+\pi^-}^2/4-m_{\pi}^2}$ is the
value of pion momentum and $|\vec{p}(m_{\sigma})|$ is the pion
momentum with on-shell $\sigma$ meson. In Eq. (\ref{second}), two
fitting parameters $F_\sigma$ and $F_{f_0(980)}$ are introduced.

Besides these two decay mechanisms mentioned above, there is the
third mechanism of $Y(4260) \to J/\psi \pi^+ \pi^-$ decay, i.e., the
ISPE mechanism, via which the typical diagrams depicting $Y(4260)
\to J/\psi \pi^+ \pi^-$ decay are given in Figs. \ref{Fig:ISPEDia}
(c)-(d) corresponding to the $\pi^-$ and $\pi^+$ emissions from
$Y(4260)$, respectively. Through the ISPE mechanism, $Y(4260)$ first
decays into a pair of intermediate charmed and anti-charmed mesons
with initial-single-pion-emission. Due to the continuous energy distribution of this pion, the intermediate charmed and anticharmed meson pair is of low momentum, which makes
these intermediate mesons easily interact with each other to transform into $J/\psi$ and $\pi$.
The diagram (c) in Fig. \ref{Fig:ISPEDia} will generate an
enhancement structure near the threshold of $D^{\ast} \bar D^{(\ast)}$ in
the $J/\psi \pi^+$ invariant mass spectrum and a similar enhancement
structure appears in the $J/\psi \pi^-$ invariant mass spectrum
resulting from diagram 1 (d). The latter enhancement 
can produce a bump when we project it into the Dalitz plot with
$J/\psi \pi^+$ invariant mass distribution in abscissa, which we call the reflection
in the $J/\psi \pi^+$ invariant mass spectrum.

The detailed calculations of $Y(4260)\to \pi^+\pi^- J/\psi$ via the
ISPE mechanism are presented in  Ref. \cite{Chen:2011xk}, which
shows that there are charged enhancement structures near
$D\bar{D}^*$ and $D^*\bar{D}^*$ thresholds. After performing the
loop integrals, the decay amplitudes via the intermediate
$D\bar{D}^*+h.c.$ and $D^*\bar{D}^*$ can be parametrized as
\begin{eqnarray}
&&\mathcal{M}_{\mathrm{ISPE}}^{D\bar{D}}[Y(4260)\to \pi^+\pi^-
J/\psi]\nonumber \\&&=g_{Y(4260)D D\pi}\epsilon_{Y(4260)}^{\mu}
\epsilon_{J/\psi}^\nu \big[ A_0  g_{\mu \nu} + \big(A_1 p_{1 \mu}
p_{1 \nu} \nonumber\\&&\quad  +A_2 p_{1 \mu} p_{2 \nu} +A_3  p_{2
\mu} p_{1 \nu}+ A_4 p_{2 \mu} p_{2 \nu}\big)\big],\\
&&\mathcal{M}_{\mathrm{ISPE}}^{D\bar{D}^*}[Y(4260)\to \pi^+\pi^-
J/\psi]\nonumber \\&&=g_{Y(4260)D^*D\pi}\epsilon_{Y(4260)}^{\mu}
\epsilon_{J/\psi}^\nu \big[ B_0  g_{\mu \nu} + \big(B_1 p_{1 \mu}
p_{1 \nu} \nonumber\\&&\quad  +B_2 p_{1 \mu} p_{2 \nu} +B_3  p_{2
\mu} p_{1 \nu}+ B_4 p_{2 \mu} p_{2 \nu}\big)\big],
\\
&&\mathcal{M}_{\mathrm{ISPE}}^{D^*\bar{D}^*}  [Y(4260)\to \pi^+\pi^-
J/\psi]\nonumber \\&&=g_{Y(4260)D^*D^*\pi}\epsilon_{Y(4260)}^{\mu}
\epsilon_{J/\psi}^\nu \big[ C_0  g_{\mu \nu} + \big(C_1 p_{1 \mu}
p_{1 \nu}\nonumber\\&&\quad + C_2  p_{1 \mu} p_{2 \nu} +C_3 p_{2
\mu} p_{1 \nu}+ C_4  p_{2 \mu} p_{2 \nu}\big)\big],
\end{eqnarray}
respectively, where the coefficients $A_i$, $B_i$, and $C_i\
\left(i=0-4\right)$, in front of the different Lorentz structures
can be evaluated by the loop integrals in the ISPE mechanism (see
Ref. \cite{Chen:2011xk} for more details). As indicated in Ref.
\cite{Chen:2011xk}, the coupling constants of $Y(4260)$, interacting
with ${D^{(\ast)} D^{(\ast)} \pi}$, $g_{Y(4260)D D\pi}$,
$g_{Y(4260)D^*D\pi}$ and $g_{Y(4260)D^*D^*\pi}$, are unknown and
become fitting parameters.

The total decay amplitude
of $Y(4260) \to \pi^+ \pi^- J/\psi$ is the sum over the subamplitudes, i.e.,
\begin{eqnarray}
&&\mathcal{M}^{\mathrm{Total}}[Y(4260) \to \pi^+ \pi^-
J/\psi]\nonumber\\ &&= \mathcal{M}_{\mathrm{Direct}} +
e^{i\phi_\sigma} \mathcal{M}_{\sigma} + e^{i\phi_{f_0(980)}}
\mathcal{M}_{f_0(980)}\nonumber\\ &&\quad+ e^{i\phi_{\mathrm{ISPE}}}
\left(\mathcal{M}_{\mathrm{ISPE}}^{D\bar{D}}
+\mathcal{M}_{\mathrm{ISPE}}^{D\bar{D}^*}
+\mathcal{M}_{\mathrm{ISPE}}^{D^\ast \bar{D}^\ast} \right),
\end{eqnarray}
where three phase angles $\phi_\sigma$, $\phi_{f_0(980)}$ and $\phi_{\mathrm{ISPE}}$ are introduced.

As discussed above, in our scenario we introduce ten free fitting
parameters as
\begin{eqnarray}
&&F, \kappa, \,F_{\sigma},\, F_{f_0(980)},\,\phi_\sigma,
\,\phi_{f_0(980)},\, \phi_{\mathrm{ISPE}}\nonumber\\& & g_{Y(4260)
DD\pi}, g_{Y(4260)D^\ast D\pi},\, g_{Y(4260) D^\ast D^\ast \pi},
\nonumber
\end{eqnarray}
which are determined by fitting to the experimental data.

With the above preparation, the differential decay width of
$Y(4260) \to \pi^+ \pi^- J/\psi$ is
\begin{eqnarray}
&&d\Gamma\left(Y(4260)\to \pi^+\pi^-
J/\psi\right)\nonumber\\&&=\frac{1}{3}\frac{1}{(2\pi)^3}\frac{1}{32m^3_{Y(4260)}}\big|\mathcal{M}^{\mathrm{Total}}\big|^2
d{m^2_{J/\psi\pi}}d{m^2_{\pi\pi}},\label{total}
\end{eqnarray}
where $m_{J/\psi\pi}^2=(p_2+p_3)^2$ and $m_{\pi\pi}^2=(p_1+p_2)^2$.
Using Eq. (\ref{total}), we can obtain the theoretical distributions
of the $J/\psi \pi^\pm$ and $\pi^+\pi^-$ invariant mass spectra,
which will be applied to fitting to the BESIII and Belle data
\cite{BESnew, Liu:2013dau}. Additionally, the resonance parameters
involving in our calculation are listed in Table \ref{rp}.

\begin{figure*}[htb]
\centering%
\begin{tabular}{ccc}
\scalebox{0.95}{\includegraphics{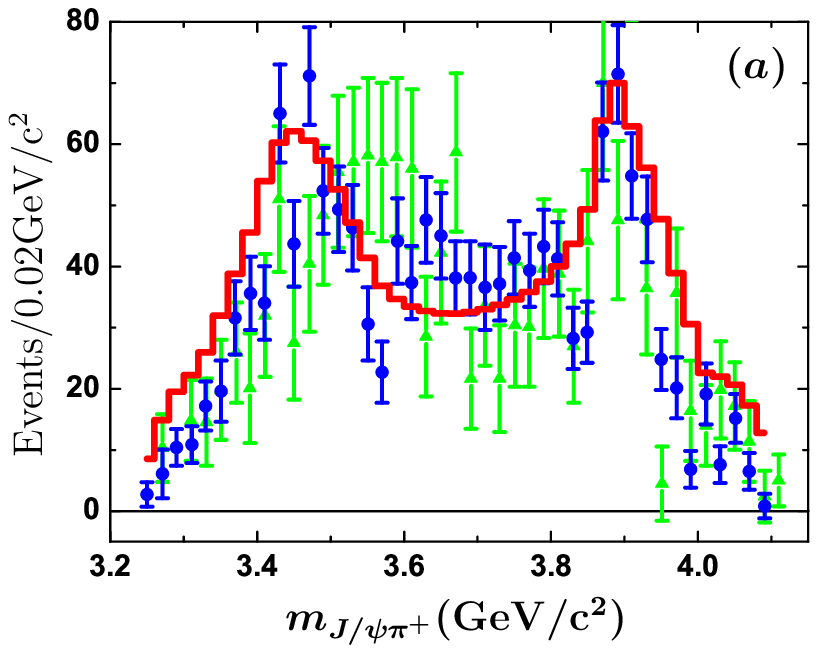}} & \hspace{5mm} &%
\scalebox{0.95}{\includegraphics{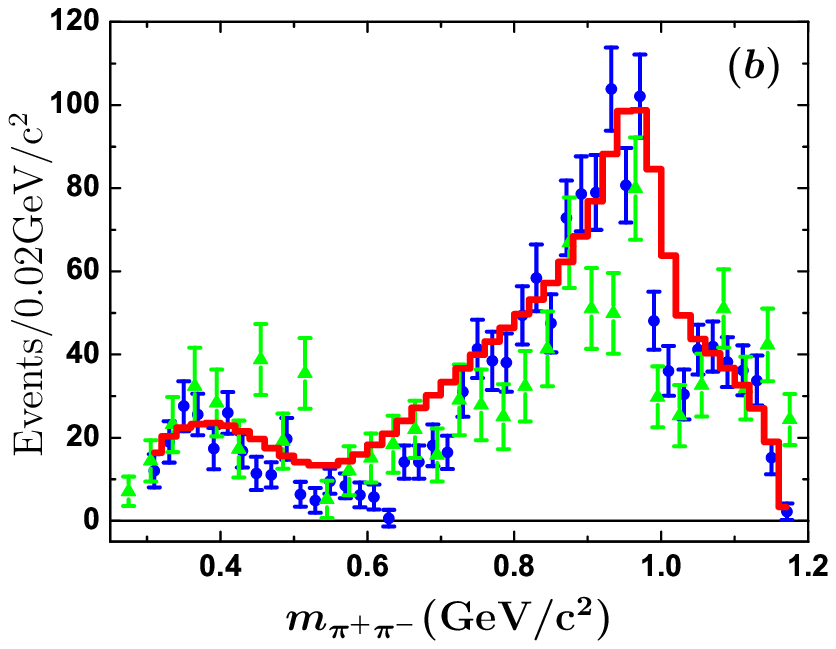}}%
\end{tabular}
\caption{(color online). The distributions of the $J/\psi\pi^+$ and
$\pi^+\pi^-$ invariant mass spectra of  $Y(4260) \to \pi^+ \pi^-
J/\psi$. The blue dots and green triangles with error bars are the
experimental data given by BESIII \cite{BESnew} and Belle
\cite{Liu:2013dau}, respectively. The red histograms are our results
considering contributions of the ISPE mechanism to the
$Y(4260) \to \pi^+ \pi^- J/\psi$ decay. \label{Fig:Fit}}
\end{figure*}

\renewcommand{\arraystretch}{1.6}
\begin{table}[htb]
\caption{The resonance parameters (in units of GeV) used in the present
work. The resonance parameters of $\sigma(600)$ are taken from Ref.
\cite{Muramatsu:2002jp} and the other resonance parameters are
from PDG \cite{Beringer:1900zz}.\label{rp}}
\begin{tabular}{cccccc}
 \toprule[1pt]
Parameter& Value&Parameter& Value&Parameter& Value\\
\midrule[1pt]
$m_{Y(4260)}        $  &  4.263  &          %
$m_{J/\psi}         $  &  3.097  &          %
$m_{\pi^\pm}        $  &  0.139     \\      %
$m_\sigma           $  &  0.513  &          %
$\Gamma_\sigma      $  &  0.335  & &\\      %
$m_{f_0(980)}       $  &  0.980  &          %
$\Gamma_{f_0(980)}  $  &  0.070  & &\\      %
\bottomrule[1pt]
\end{tabular}
\end{table}

\renewcommand{\arraystretch}{1.5}
\begin{table}[htb]
\caption{The obtained values of the parameters for the optimum fit to the BESIII and Belle data of $Y(4260) \to \pi^+ \pi^- J/\psi$ decay \cite{BESnew}.\label{T1} }
\begin{tabular}{cccc}
\toprule[1pt]
Parameter &  Value & Parameter &  Value \\
\midrule[1pt]
 $F                       $         &  $ -21.56  $  &                   %
 $\kappa                  $         &  $   0.92  $  \\                  %
 $f_\sigma                $         &  $ 140.00\, \mathrm{GeV}^2$  &     %
 $\phi_\sigma             $         &  $   2.45 $      \\              %
 $f_{f_0(980)}            $         &  $  60.00  \,\mathrm{GeV}^2$  &     %
 $\phi_{f_0(980)}         $         &  $   0.25 $  \\                  %
 $g_{Y(4260) D D \pi}     $         &  $ 25.37 \ \mathrm{GeV}^{-3}$ & %
 $g_{Y(4260) D^\ast D \pi}$         &  $  32.50 $  \\                   %
 $g_{Y(4260) D^\ast D^\ast \pi}$    &  $  -4.70 \,\mathrm{GeV}^{-1}$  & %
 $\phi_{\mathrm{ISPE}}    $         &  $  -0.26 $         \\        %
\bottomrule[1pt]
\end{tabular}
\end{table}

At present, both BESIII and Belle have not only given the $J/\psi
\pi^\pm$ invariant mass distributions but have also provided the
$\pi^+\pi^-$ invariant mass spectrum \cite{BESnew, Liu:2013dau}.
This information provides us with a good chance to test the ISPE
mechanism. Having Eq.~(\ref{total}), we try to reproduce the data of
$Y(4260) \to \pi^+ \pi^- J/\psi$ from BESIII and Belle, which is
helpful to test our ISPE mechanism. The interference effects of the
ISPE mechanism with the other two decay mechanisms are also included in
the present work. The sideband contributions are taken as a
background of the measurements, which are subtracted in
Fig.~\ref{Fig:Fit}. The Belle data is normalized to the BESIII data
by multiplying by a factor, which is the ratio of total events of
BESIII and Belle for corresponding bins, and for dipion invariant
mass distributions an additional factor 2/3 is also multiplied with
Belle data, which is due to a different bin width between Belle and
BESIII data.

Using Eq.~(\ref{total}), we fit the data from the BESIII and Belle
collaborations and try to reproduce the enhancement in the $J/\psi
\pi^+$ invariant mass spectrum, whose peak is reported as
$Z_c(3900)$. The two red histograms in Fig.~\ref{Fig:Fit} are the
results of our model with the ISPE mechanism. To obtain these
histograms, we take the optimum values listed in Table \ref{T1}.
Comparing our theoretical results with the data from the BESIII and
Belle collaborations, we notice that the $d\Gamma(Y(4260)\to
\pi^+\pi^- J/\psi)/dm_{J/\psi \pi^\pm}$ distribution agrees well 
with the $J/\psi \pi^\pm$ invariant mass spectrum. Here, two peaks,
the $Z_c(3900)$ structure and its reflection, can be nicely
reproduced, where our fitting results indeed reflect the fact that
$Z_c(3900)$ is much sharper than its reflection. In addition, with
the same fitting parameters, the theoretical distribution of
$d\Gamma(Y(4260)\to \pi^+\pi^- J/\psi)/dm_{\pi^+ \pi^-}$ is also
given and reproduces well the experimental data of the $\pi^+\pi^-$
invariant mass spectrum again.

In summary, the charged charmoniumlike structure $Z_c(3900)$ has
been newly observed and reported by the BESIII Collaboration
\cite{BESnew} and confirmed by Belle \cite{Liu:2013dau} and CLEO-c
\cite{Xiao:2013iha}. After this new observation, different
theoretical groups have given different explanations for $Z_c(3900)$
\cite{Wang:2013cya,Guo:2013sya,Faccini:2013lda,
Voloshin:2013dpa,Karliner:2013dqa,Mahajan:2013qja,
Cui:2013yva,Wilbring:2013cha}. Among these explanations, the
discussions on whether the structure of $Z_c(3900)$ is due to an
exotic state or not are very popular, where the exotic state
explanation of $Z_c(3900)$ includes a molecular state composed of $D$
and $\bar{D}^*$ mesons or a tetraquark state.

As mentioned in the BESIII experimental paper \cite{BESnew}, there
existed theoretical predictions of charged charmoniumlike
structures near the $D\bar{D}^*$ threshold
\cite{Chen:2011xk,Chen:2012yr,Sun:2011uh} before the observation of
$Z_c(3900)$. In Ref. \cite{Chen:2011xk}, the ISPE mechanism was
applied to study $Y(4260)\to \pi^+\pi^-
J/\psi,\,\pi^+\pi^-\psi^\prime,\,\pi^+\pi^- h_c$ decays, where two
of the authors of this paper predicted the interesting phenomena of
charged $D\bar{D}^*$ and $D^*\bar{D}^*$ structures existing in the
$J/\psi\pi^\pm$, $\psi^\prime \pi^\pm$ and $h_c\pi^\pm$ invariant
mass spectra. The BESIII observation of $Z_c(3900)$ has inspired us
to study the $Y(4260)\to \pi^+\pi^- J/\psi$ decay process by the
ISPE mechanism again together with other diagrams shown in Fig.
\ref{Fig:ISPEDia}, where we are able to fit our model completely, 
including a tree diagram as well as relative phases to the
experimental data in our scenario, since BESIII now provides abundant
information for the distribution of the $J/\psi\pi^\pm$ and
$\pi^+\pi^-$ invariant mass spectra.

As shown in Fig. \ref{Fig:Fit}, the $Z_c(3900)$ structure has been
remarkably reproduced under the ISPE mechanism that includes the
interference effects. This fact affirmatively answers whether
$Z_c(3900)$ corresponds to the predicted charged charmoniumlike
structure near $D\bar{D}^*$ threshold via the ISPE mechanism in Ref.
\cite{Chen:2011xk}.

Besides the prediction and/or reproduction relevant to $Z_c(3900)$
under the ISPE mechanism, there are other abundant predictions for
the features of charged enhancement structures in the hidden-charm
dipion decays of higher charmonia \cite{Chen:2011xk}, the
hidden-bottom dipion decays of $\Upsilon(11020)$ \cite{Chen:2011pu},
and the hidden-strange dipion decays of $Y(2175)$
\cite{Chen:2011cj}. We hope that future experiments will carry out
searchs for these novel phenomena.

Given the experimental data, the present work has provided an
important approach to test the ISPE mechanism. With more
experimental observations, more phenomenological and theoretical
efforts should be made by combining further experimental data, which
will finally clarify what is the underlying mechanism behind the
observed $Z_c(3900)$ structure and other exotic structures.

\vfil
\section*{Acknowledgements}

This project is supported by the National Natural Science
Foundation of China under Grants No. 11222547, No. 11175073, No.
11005129 and No. 11035006, the Ministry of Education of China
(FANEDD under Grant No. 200924, SRFDP under Grant No.
2012021111000, and NCET), the Fok Ying Tung Education Foundation
(No. 131006), and the West Doctoral Project of Chinese Academy of
Sciences.

\end{document}